\documentclass[preprint,showpacs,showkeys,preprintnumbers,amsmath,amssymb]{revtex4}

\usepackage{graphicx}
\usepackage{dcolumn}
\usepackage{bm}

\begin{document}

\title{Electronic transport calculations for rough interfaces in Al, Cu, Ag, and Au}

\author{M.\ M.\ Fadlallah$^{1,2}$, C.\ Schuster$^1$, U.\ Schwingenschl\"ogl$^{1}$, T.\ Wunderlich$^1$, and S.\ Sanvito$^3$}
\affiliation{
$^1$Institut f\"ur Physik, Universit\"at Augsburg, 86135 Augsburg, Germany\\
$^2$Physics Department, Faculty of Science, Benha University, Benha, Egypt\\
$^3$School of Physics and CRANN, Trinity College, Dublin 2, Ireland}
\date{\today}

\begin{abstract}

We present results of electronic structure and transport calculations for metallic 
interfaces, based on density functional theory and the non-equilibrium Green's functions method.
Starting from the electronic structure of smooth Al, Cu, Ag, and Au interfaces, we study the
effects of different kinds of interface roughness on the transmission coefficient and the
I-V characteristic. In particular, we compare prototypical interface distortions, 
including vacancies and metallic impurities. 
\end{abstract}

\pacs{73.20.-r, 73.40.-c, 73.40.Jn}
\keywords{Density functional theory, non-equilibrium transport, metallic interface, impurity} 

\maketitle

\section{Introduction}

Electronic devices have been more and more reduced in size over the last decades. Today their typical scale is
some ten nanometers, so that quantum mechanical effects cannot be neglected anymore.
Clearly, their functionality depends crucially on the transport characteristics of interfaces.
For a better understanding of such devices a quantum theory of transport is required. Due to
recent progress in methods of placing atoms or molecules between macroscopic electrodes, many studies 
of the electronic transport have been possible for single atoms, molecules, and nanowires \cite{Ga,Ze,Fin,Cor,Ha,Rocha}.
In addition, interface properties, like interface charging, band bending, or the contact resistivity, have been
investigated extensively.

Interfaces can give rise to new features not present in
any of the components, like the creation of conduction states between two bulk insulators \cite{thiel06,US_LAO}.
Another important example is the charge redistribution at metal-superconductor interfaces, which has a major
impact on electronic applications \cite{US_apl}. Electronic phenomena occuring at semiconductor interfaces have
been reviewed comprehensively by M\"unch \cite{Book4}. Moreover, the induced charge density in organic/inorganic devices
was subject of thorough studies \cite{scheffler,bagus05,US_benzen}, and the orientation dependent transparency
of simple metals, like Al and Ag, has been analyzed \cite{xu06}. However, the transport properties of simple
metal interfaces have not been investigated in detail. In particular, interface distortions, like atomic scale
roughness, vacancies, and impurities, are of considerable practical interest, while theoretical results on this issues
are rare due to the structural complexity. For example, the influence of disorder on the interface resistance in
Au/Ag multilayers has been studied using large lateral supercells \cite{xia01}.

Several methods, based on electronic structure calculations, have been developed to address the problem of 
transmission at interfaces and nanocontacts. To describe a nanoscaled device connected to leads, an (open) non-periodic
system must be considered. Therefore, standard repeated-slab or cluster models, which have sucessfully been applied to
static interfaces in equilibrium, fail in this case.
A theoretical description of a nanometer scale junction in a strong electric field was presented by Hirose and
Tsukada \cite{Hi,Lang}. In addition, a tight-binding formulation was applied to metallic nanocontacts \cite{chico,cuevas}. 
The layer extension of the Korringa--Kohn--Rostoker method allows to describe the propagation through a layered system, and overcomes the restiction to periodic systems \cite{Mac}. 
Another approach to the electronic transport through interfaces is given by the embedded Green's function scheme \cite{Cra,Wo}.

The generalization of density functional theory (DFT) to time dependent potentials \cite{runge} allows to study the propagation
of the electronic states in a rigorous manner \cite{kurth}. 
However, most computational methods rely on a combination of DFT and a scattering theory at the non-equilibrium Green's functions level,
based on the Landauer--B\"uttiker scheme. In particular, our present study employs the SMEAGOL program package \cite{Roch1,Roch2},
a flexible and efficient implementation which consists of a direct summation of both open and closed scattering channels
together with a regularization procedure for the Hamiltonian. As a consequence, materials with complex electronic structures can be tackled.

In this paper, the electronic structure and the transmission coefficient for transport through different distorted
interfaces of simple metals are determined. The I-V characteristic is likewise addressed.
In Sec.\ II we give details about the calculational method and the structural setup. In Sec.\ III we discuss the results
of our equilibrium electronic structure calculations, in particular, the density of states (DOS) and the transmission coefficient $T(E,V=0)$.
Finally, in Sec.\ IV the non-equilibrium properties of the systems are investigated for a finite voltage $V\neq0$.

\section{Computational details and structural setup}

Among the packages using a combination of DFT and the non-equilibrium Green's functions method for calculating transport
properties, we choose SMEAGOL \cite{Roch1,Roch2} for our investigations. SMEAGOL is based on the DFT package SIESTA \cite{Sol}
which provides the required basis set of local atomic orbitals. The surface Green's functions are calculated using
semi-analytical expressions, which results in a great improvement with respect to recursive methods \cite{rungger08}.

The McDCal code \cite{Tay}, for example, likewise relies on an atomic orbital basis \cite{Ord} but solves the
electrostatic problem by a real-space multi-grid approach \cite{Bran}. On the other hand, the GECM transport code \cite{Pala,Lo}
simplifies the treatment of the leads by introducing a tight-binding Bethe lattice \cite{Joa}.
The TranSIESTA code \cite{Bra} is an extension of SIESTA, solving the electrostatic problem in momentum space
and calculating the surface Green's functions by direct integration. 

SMEAGOL, in particular, is designed for materials with a complicated electronic structure and with spin polarization.
To apply the Landauer--B\"uttiker formalism, first the self-energies of the left (L) and right (R) lead are calculated.
Then the leads are connected to a central region of interest (molecule, nano-contact, interface), where the
whole device is in equilibrium at the beginning. In equilibrium, the transmission coefficient is given by the retarded Green's function
$G_M$ of the central (M) region and the lead self energies $\Sigma_{L/R}$, with $\Gamma_{L/R}:={\rm i}[\Sigma_{L/R}(E)-\Sigma_{L/R}^{\dagger}(E)]$,
\begin{displaymath}
T(E,V=0)={\rm Tr}[\Gamma_L G_{M}^{\dagger}\Gamma_R G_{M}].
\end{displaymath}
Applying an external voltage $V$, non-equilibrium Green's functions have to be considered; the charge density
can be calculated from the ``lesser'' Green's function
\begin{eqnarray*}
G^<_M(E)&=&iG_M(E)[\Gamma_L(E-eV/2)\cdot f(E-(\mu-eV/2))\\
&&+\Gamma_R(E+eV/2)\cdot f(E-(\mu+eV/2))]G^\dagger_M(E)
\end{eqnarray*}
with the Fermi function $f$ and the chemical potential $\mu$. The current then can be written as
\begin{eqnarray*}
I(V)&=&{e\over{h}}\int dE\;T(E,V)\cdot\\
&&[f(E-(\mu-eV/2))-f(E-(\mu+eV/2))].
\end{eqnarray*}

We wish to study internal interfaces of specific materials. In particular, we are interested in smooth and distorted interfaces of prototypical
metals and their transmission coefficients. We choose Al as a canonical example of an $sp$-hybride metal with $3s^23p^1$ valence state,
which is well described by the free electron model. Its DOS is almost perfectly proportional to the square root of the energy
\cite{Book1}. In addition, we address the noble metal Au, which is commonly used in transport experiments,
and the narrow $d$ band transition metal Cu as a typical impurity compound. Nobel metals like Cu ($3d^{10}4s^{1}$), Ag ($4d^{10}5s^{1}$),
and Au ($5d^{10}6s^{1}$) have a very similar valence electronic structure: fully occupied $d$ states at about 2 eV below
the Fermi level and a single $s$ valence electron (with substantial $sd$-hybridization). Moreover, these metals are
well suitable for a comparative study since they all crystallize in a fcc structure. We address the [001] transport
direction for simplicity.

SIESTA uses norm-conserving pseudopotentials in the fully non-local form (Kleinman-Bylander \cite{kleinman}). For the noble metals, we apply pseudopotentials
including $d$ valence states (and $f$ states for Au). Moreover, a double zeta basis set and the generalized gradient approximation for the
exchange correlation potential are used. The parameters of the SMEAGOL calculation are the following: The lead comprises
two unit cells to form a principal layer. Furthermore, the scattering region
consists of six unit cells in each case. For the lead calculation we make use of a mesh of $15\times15\times100$ $k$-points, while the
$k$-mesh in the transport calculation is $10\times10\times1$. Finally, to determine
the density matrix, we choose up to 50 energy points on the semi-circle in the complex plane,
up to 100 energy points along the line in the complex plane, and up to 50 poles in the Fermi distribution.

\begin{figure}[b]
\includegraphics[width=0.35\textwidth,clip]{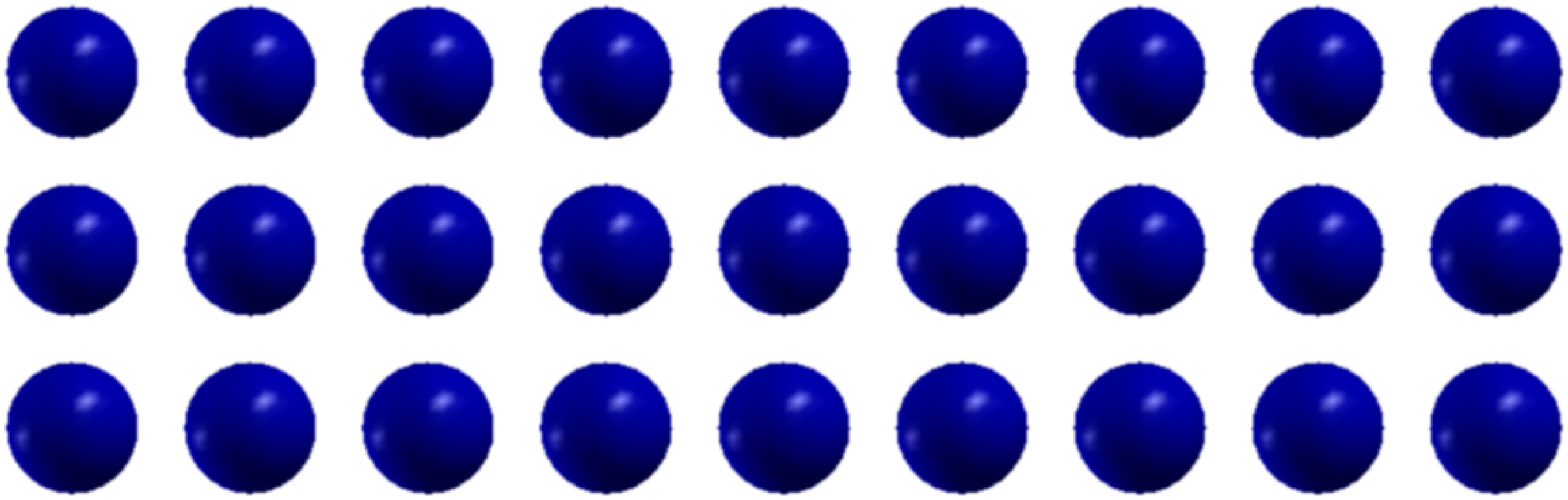}\\[5mm]
\includegraphics[width=0.35\textwidth,clip]{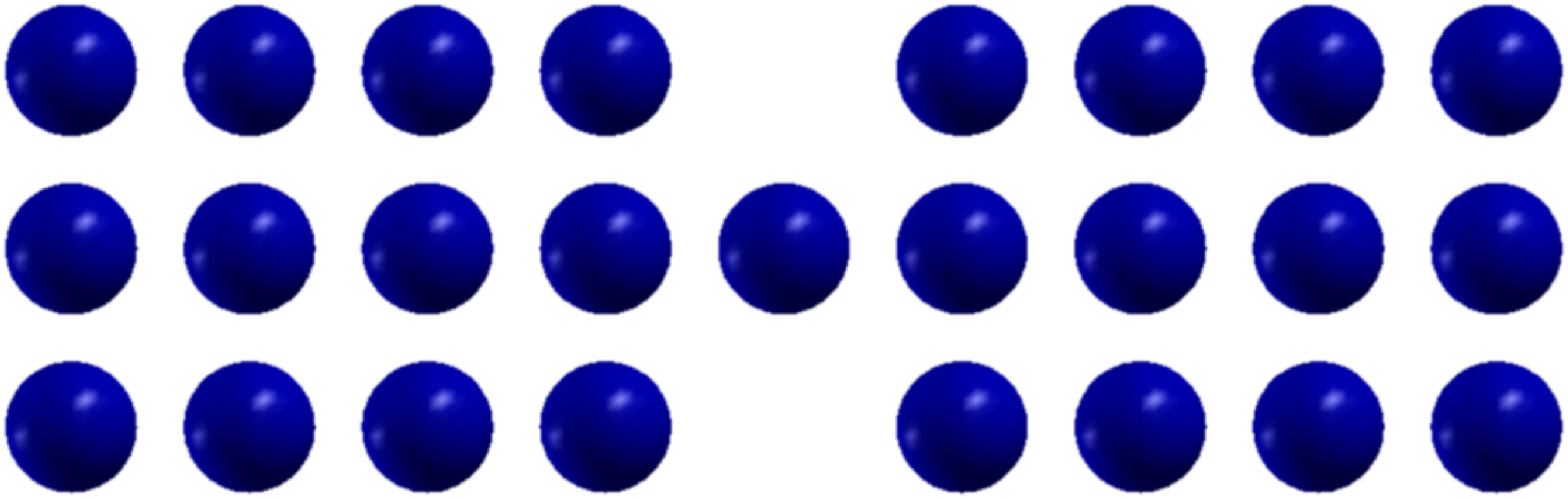}\\[5mm]
\includegraphics[width=0.35\textwidth,clip]{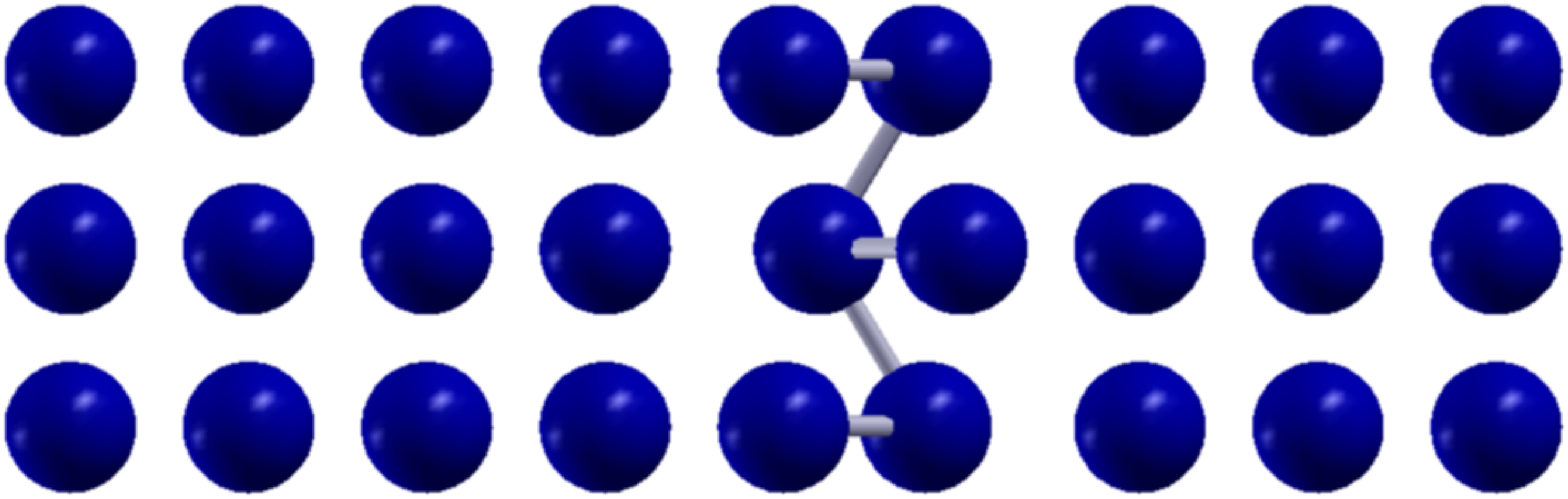}\\[5mm]
\includegraphics[width=0.35\textwidth,clip]{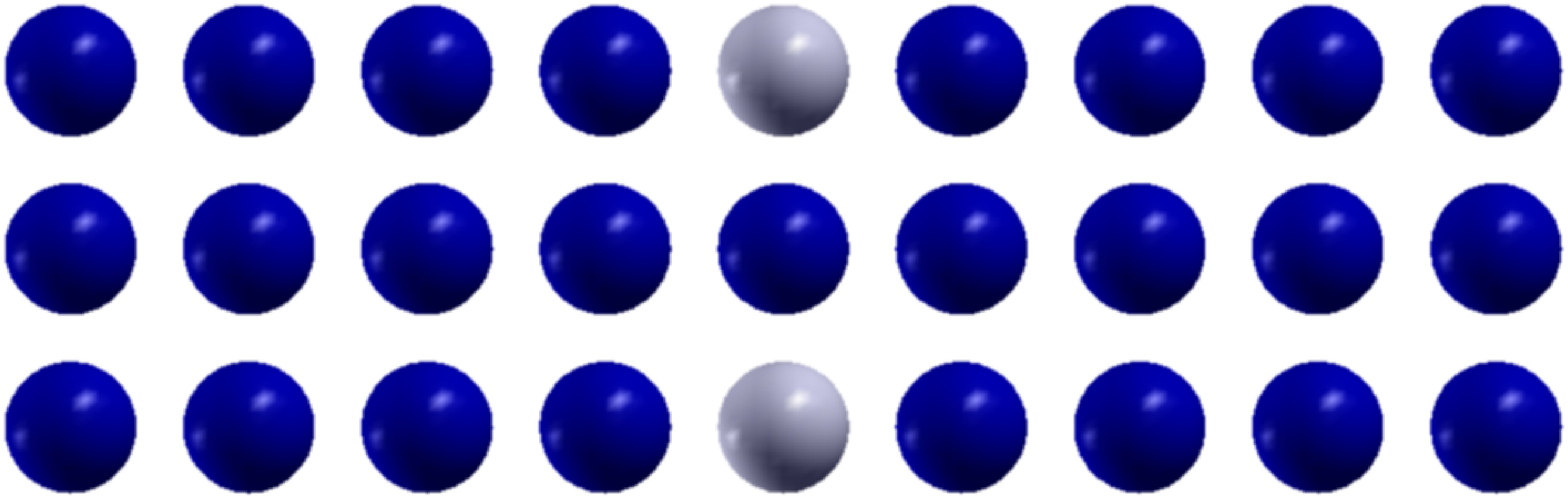}\\[5mm]
\includegraphics[width=0.35\textwidth,clip]{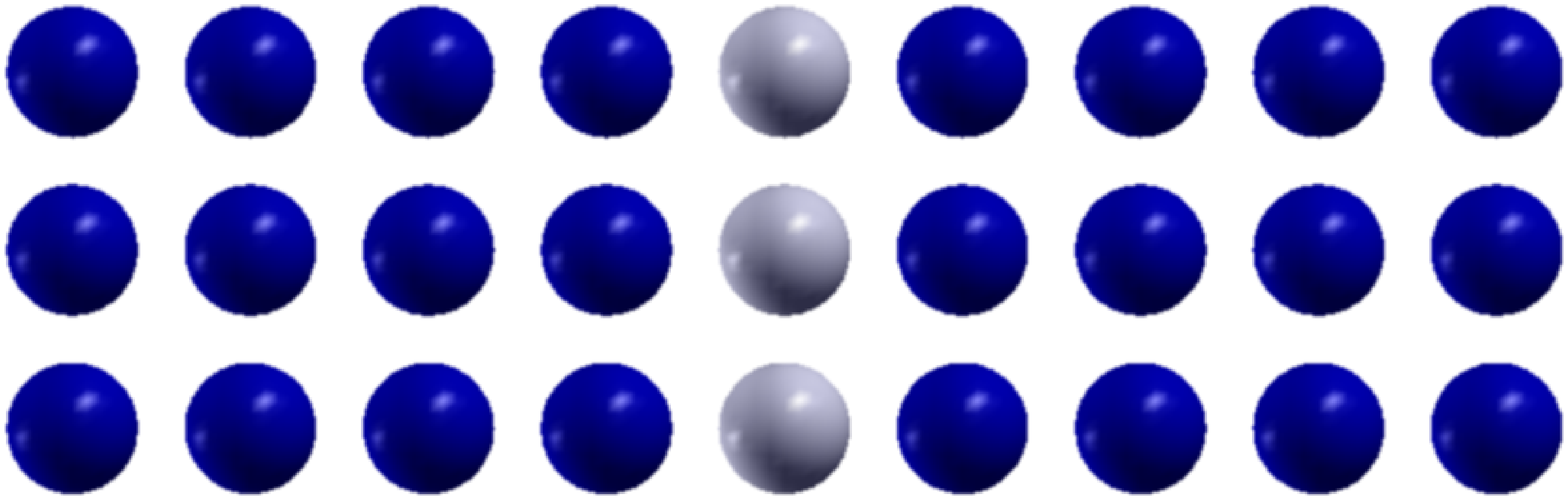}\\[5mm]
\caption{Structures under investigation (from top to bottom): smooth interface, interface with vacancy, 
buckled interface, interface with impurity, and metallic interlayer.}
\label{fig1}
\end{figure}

Being interested in the transport along fcc [001] direction,
our discussion will focus on these interface distortions: Smooth Al and Au interfaces are compared to interfaces with vacancy
and impurity sites, a buckled interface plane, and an impurity interlayer. The structures are displayed in Fig \ref{fig1}.
The vacancy configuration, which we study for Al and Au as well as for Cu and Ag, is created by removing each second
interface atom. It therefore models a series of mono-atomic contacts. Inserting Cu atoms into the vacancies of the Au host
leads to an impurity structure, where the impurity electronic states are closely related to those of the host.
When simulating a buckled interface, we shift neighbouring interface atoms zigzag-like by 10\% of the fcc lattice constant along
the [001] direction. Finally, the interlayer configuration consists of a mono-atomic Cu layer incorporated in the Au host.
In order to highlight the effects of our principal distortions, further structural relaxation effects are not taken
into account. This is well justified because previous studies of Al nanocontacts have indicated only minor alterations
of the bond lengths in the contact region \cite{US_Al_1}.

\section{Equilibrium properties}

\begin{figure*}[t]
\includegraphics[width=0.7\textwidth]{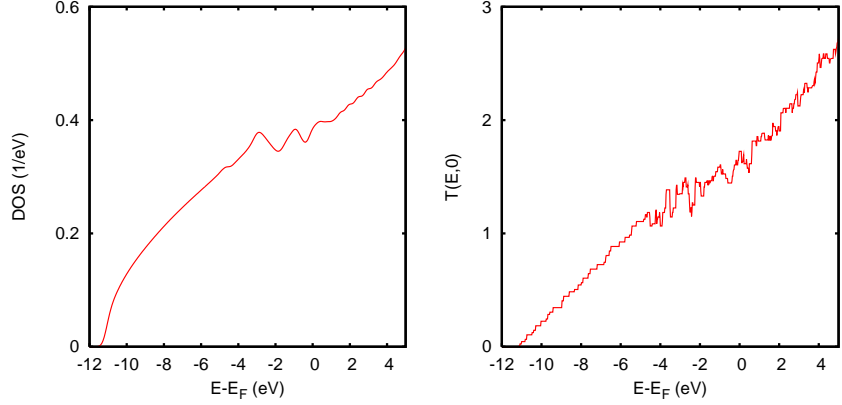}
\caption{Al DOS (left hand side) and transmission coefficient (right hand side) of the smooth Al interface.}
\label{fig2}
\end{figure*}

\begin{figure*}[t]
\includegraphics[width=0.7\textwidth,clip]{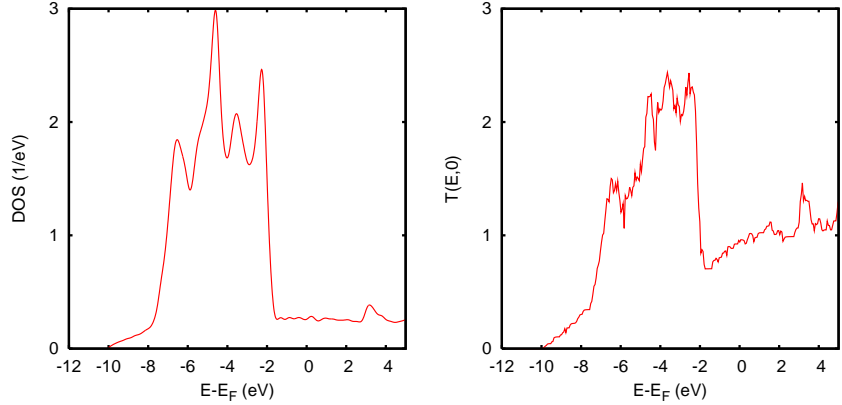}
\caption{Au DOS (left hand side) and transmission coefficient (right hand side) of the smooth Au interface.}
\label{fig3}
\end{figure*}

Since SMEAGOL originally has been designed for investigating mono-atomic contacts, we first study whether it likewise has the capacity
of dealing with our extended systems. To this end, we address smooth Al and Au interfaces, for which we show the DOS and transmission
coefficient $T(E,V=0)$ in Figs.\ \ref{fig2} and \ref{fig3}, respectively. $T(E,V=0)$ is normalized by the number of transverse k-points. We find that  
the shape of $T(E,V=0)$ is closely related to the DOS shape, both for the $sp$-hybride system and the noble metal system.
For the smooth Al interface, onset of transmission is found when electronic states are available near $-11.5$ eV, see Fig.\ \ref{fig2}. 
As to be expected for the nearly free electrons of the Al system, the transmission coefficient on the right hand side of
Fig.\ \ref{fig2} reveals an almost ideal linear behaviour up to $-5$ eV and only minor deviations for higher energies.
Reflecting the DOS shape, the transmission through the smooth Au interface starts near $-10$ eV and  
shows high values between $-8$ eV and $-2$ eV due to the broad Au 5$d$ bands, see Fig.\ \ref{fig3}. 
Importantly, in the vicinity of the Fermi energy ($E_F$) the transmission coefficient $T(E,V=0)$ grows, in a good
approximation, linearly with increasing energy, reflecting the delocalized nature of the corresponding states.
The quantitative evaluation of the transmission coefficient yields for the conductance $G=G_0\cdot T(E_F,V=0)$,
with $G_0=2e^2/h$, values of about 1 $G_0$ and 1.5 $G_0$ for the Au and Al interface, respectively.


\begin{figure*}[t]
\includegraphics[width=0.7\textwidth,clip]{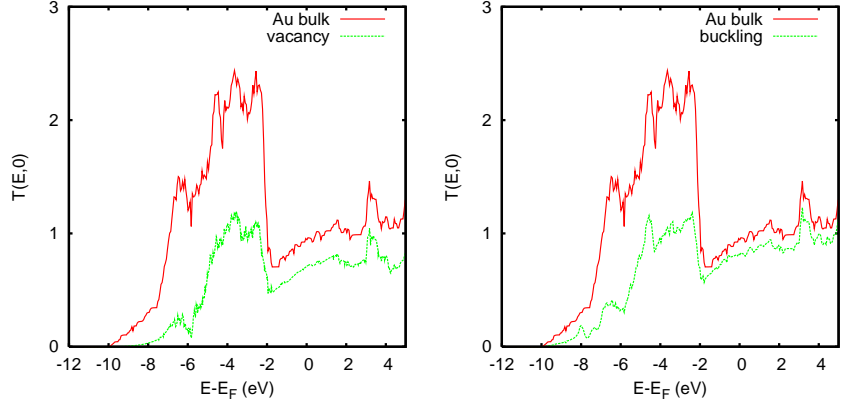}
\caption{Transmission coefficient of the Au interface with vacancy (left hand side) and
the buckled interface (right hand side), compared to the smooth Au interface.}
\label{fig4}
\end{figure*}

Turning to rough Au interfaces, we now address the distortions introduced previously.
Fig.\ \ref{fig4} indicates that insertion of vacancy sites leads to a clear reduction
of the transmission in the full energy range. At the transmission maximum around $-4$ eV
the amplitude is reduced by 55\% as compared to the smooth interface, i.e., the bulk
system. For $E>-2$ eV the reduction is less pronounced, amounting to 25\%. An interface
buckling has a similar effect, as illustrated on the right hand side of Fig.\ \ref{fig4}.
In particular, almost the same reduction of the transmission coefficient is observed
below $-2$ eV, whereas in the energy range from $-2$ eV to $5$ eV it is significantly smaller,
amounting to 15\%. This finding can be attributed to the well-known differences in the localization of the Au electronic
states below and above $-2$ eV. For $E<-2$ eV the average group velocity is lower 
and the effective mass is higher. Therefore, the impurity scattering is enhanced. Since
$d$ bands are more sensitive to local disorder \cite{prb77} and the transmission
is almost completely due to the $d$ channel for $E<-2$ eV, it is of no surprise that the
reduction of the transmission coefficient is more pronounced in this portion of the
electronic spectrum. We finally mention that for the buckled interface the onset of
$T(E,V=0)$ already appears at $-10$ eV, while it is found at $-8$ eV for the vacancy system.

\begin{figure*}[t]
\includegraphics[width=0.7\textwidth,clip]{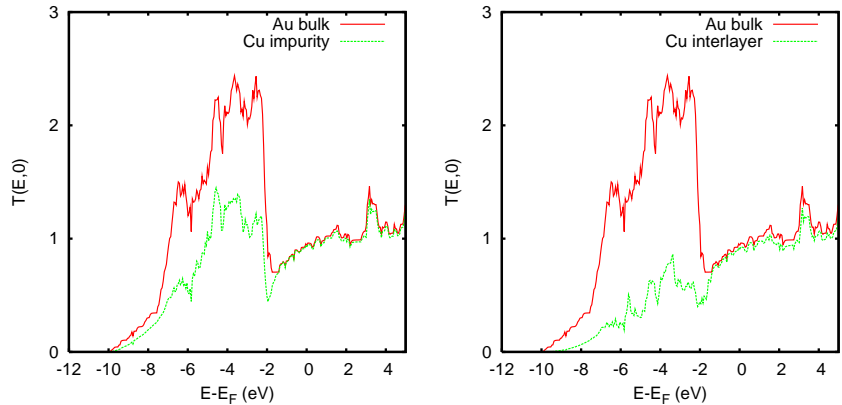}
\caption{Transmission coefficient of the Au interface with Cu impurity (left hand side) and
Cu interlayer (right hand side), compared to the smooth Au interface.}
\label{fig5}
\end{figure*}
\begin{figure*}[t]
\includegraphics[width=0.33\textwidth,clip]{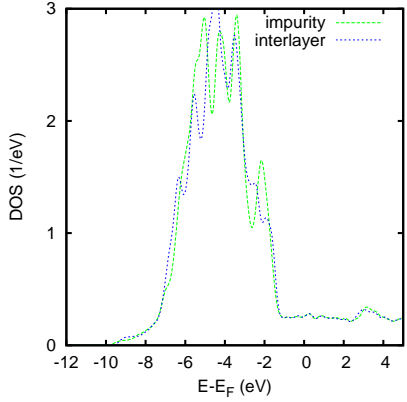}
\caption{Local Au DOS at the interfaces with Cu impurity and Cu interlayer.}
\label{fig5b}
\end{figure*}

Filling the vacancy sites with Cu atoms, the transmission increases considerably, compare
the left hand sides of Figs.\ \ref{fig4} and \ref{fig5}. Due to the stronger localization
of the Cu 3$d$ orbitals, the transmission is reduced by 40\% below  $-2$ eV, with respect
to the smooth Au interface. On the contrary, we observe no reduction for the higher energy itinerant
states. Extending the isolated impurity sites to a full impurity interlayer, the suppression
of the transmission is remarkably enhanced below $-2$ eV, see the right hand side of Fig.\
\ref{fig5}. According to the above argumentation, we expect that states above $-2$ eV are
hardly influenced. Both for the Cu impurity and the Cu interlayer the electronic structure
of neighbouring Au atoms reveals only minor modifications, compare Fig.\ \ref{fig5b} to the
left panel of Fig.\ \ref{fig3}. Even though metallic impurities result in a drastic
reduction of the transmission coefficient far below the Fermi level, the transport properties,
which depend on $T(E_F,V=0)$, are not altered. The conductance is considerable smaller for
our vacancy system than for our buckled interface, whereas impurities play a minor role.

\begin{figure*}[t]
\includegraphics[width=0.7\textwidth,clip]{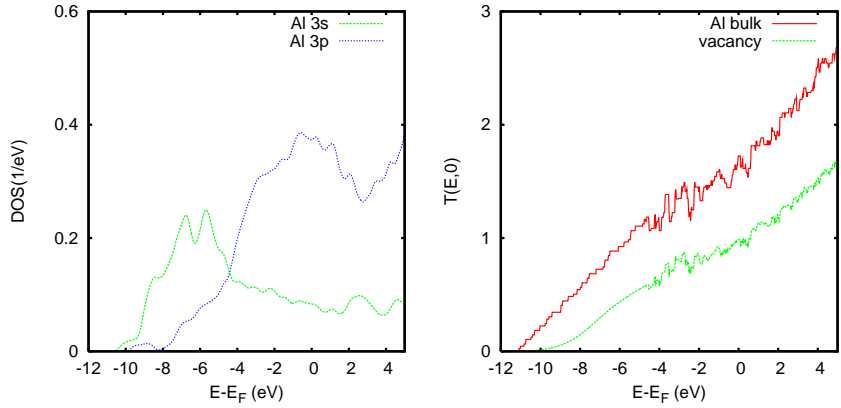}
\caption{Local Al $3s$ and $3p$ DOS (left hand side) and transmission coefficient (right hand side) of the Al
interface with vacancy. The transmission coefficient of the corresponding smooth interface is
given for comparison.}
\label{fig6}
\end{figure*}

\begin{figure*}[t]
\includegraphics[width=0.7\textwidth,clip]{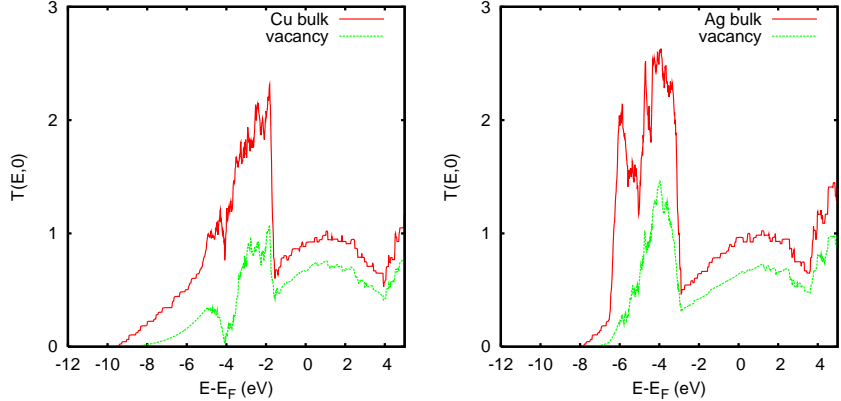}
\caption{Transmission coefficient of the Cu (left hand side) and Ag (right hand side)
interfaces with vacancy, each compared to the corresponding smooth interface.}
\label{fig6b}
\end{figure*}

Next we study the influence of vacancies in the case of the $sp$-hybride Al system, see
Fig.\ \ref{fig6}. Almost independent of $E$, we observe a substantial reduction of
$T(E,V=0)$ by 40\% with respect to the values of the smooth Al interface. Interestingly,
the linear slope of $T(E,V=0)$ below $-4$ eV is not reproduced by the vacancy calculation,
which we attribute to a strong perturbation of the Al electronic structure. The latter is
reflected by the partial Al $3s$ and $3d$ DOS curves shown in the left panel of Fig.\
\ref{fig6}, which refer to an atom next to a vacancy. The perturbation traces back to a
reduced isotropy of the chemical bonding and, therefore, a suppression of the
$sp$-hybridization \cite{US_Al_1}. For this reason, the reduction of $T(E_F,V=0)$ is much
larger than for the Au host, see the left hand side of Fig.\ \ref{fig4}. In contrast to
the Al host, the effects of vacancies in a Cu/Ag host, see Fig.\ \ref{fig6b}, are very
similar to the Au case, compare our previous discussion. Of course, the latter is expected
from the close relations between the Cu, Ag, and Au electronic structures. In particular,
we observe the broad transmission peak due to the metal $d$-states and the linear energy
dependence of $T(E,V=0)$ in the vicinity of $E_F$.

\section{Non-equilibrium properties}

In this section, we address the non-equilibrium transport properties of smooth and defective Al and Au interfaces.
Data for Cu and Ag interfaces are not presented here since they are rather similar those of Au.
Note that in general the application of a potential bias across an homogeneous metallic conductor
does not produce a potential drop, which, in contrast, might be localized at the contact region. This corresponds
to the known fact that a pure metal cannot sustain an internal electric field. Nevertheless it is interesting
to investigate how an artificially imposed potential drop affects the transport properties of both clean
and defective interfaces. This gives us an indication on how the system responds to an external electrical
perturbation. Within SMEAGOL it is possible to impose such a potential drop by simply setting the difference
between the chemical potentials of the leads to $eV$, in such a way that a potential drop of $V$ artificially
establishes across the scattering region. Note that the resulting external electric field then depends on the 
length of the scattering region itself. With this in mind we have re-calculated the transmission coefficient
as a function of energy for different imposed potential drops.

\begin{figure*}[t]
\includegraphics[width=0.7\textwidth,clip]{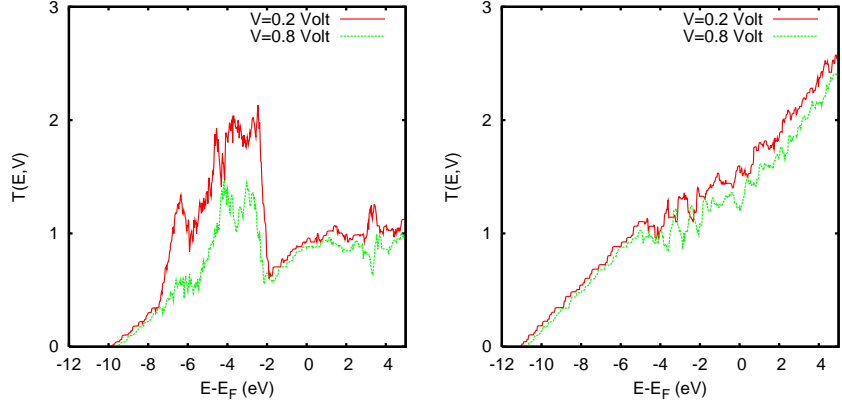}
\caption{Transmission coefficient of the smooth Au (left hand side) and Al (right hand side) interfaces at voltages of 0.2 V and 0.8 V.}
\label{fig7}
\end{figure*}

\begin{figure*}[t]
\includegraphics[width=0.7\textwidth,clip]{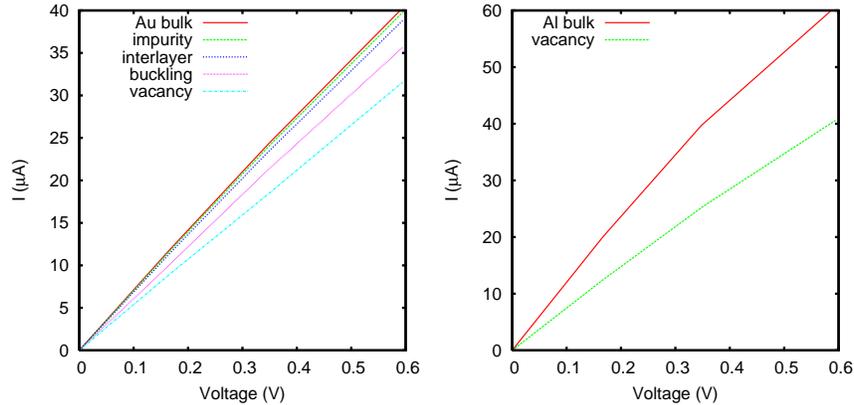} 
\caption{Left hand side: I-V characteristics for the various Au interfaces, see Fig.\ \ref{fig1}. 
Right hand side: I-V characteristics for the smooth Al interface and the Al interface with vacancy.}
\label{fig8} 
\end{figure*}

The results are shown in Fig.\ \ref{fig7}, where we use the same convention as for the $V=0$ case to plot $T(E,V\neq0)$ 
with the energy scale measured relative to the equilibrium $E_F$. Since the external potential $V$ disturbs 
the periodicity of the electronic states along the transport direction, a reduction of $T(E,V\neq0)$, as compared to 
$T(E,V=0)$, is expected. This expectation is partially confirmed in the case of the smooth Au interface (see the left hand 
side of Fig.\ \ref{fig7}), for which $T(E,V=0.8$ V) is reduced by 40\% with respect to $T(E,V=0)$ in the energy range 
corresponding to the Au $5d$ density of states and by $\approx$ 10\% well above the Fermi level. In contrast, 
the reduction of the transmission coefficient is much smaller for the smooth Al interface (see the right hand side
of Fig.\ \ref{fig7}), amounting to $\approx$ 10\% above $-4$ eV. Below this energy, we have no reduction
at all. Of course, this observation finds its origin in the free electron character of the Al electronic states below $-4$ eV,
with the same explanation applying to the Au data at the low energy edge and around $E_F$.

Since for both the Al and Au interfaces the value of the transmission coefficient at $E_F$ is
hardly modified by the external potential, little changes in the conductivity are predicted in the linear
response. The calculated $I$-$V$ characteristics obtained by artificially imposing the potential drop
are presented in Fig.\ \ref{fig8} and support this conjecture. In the same figure we also report results
for a number of selected defective interfaces, which also show almost perfectly linear $I$-$V$ characteristics. 
In general we find that the reduction of $T(E_F,V=0)$, see the discussion in Sec.\ III, nicely 
correlates with the reduction in conductance, with respect to the bulk values, extracted from these $I$-$V$ data, 
respectively of 25\% for vacancies and to 15\% for the buckled interface. As expected from the values obtained for 
$T(E_F,V=0)$, the incorporation of Cu impurities (isolated or interlayer) has only a minor effect on the 
Au conductance.

\section{Summary}

In summary, we have confirmed the capability of the SMEAGOL program package to deal with extended systems.
We have addressed the transport through smooth and distorted fcc Al, Cu, Ag, and Au interfaces, where
we particularly have identified regimes of (nearly) free electrons via the linear energy dependence of
the transmission coefficient. Investigating various kinds of distortions, we find that vacancy sites
have a huge effect on $T(E_F,V=0)$. Buckling of the interface atomic layers reduces the transmission
strongly when localized Cu, Ag, and Au $d$ states are involved. A relevant reduction is also found for (nearly)
free electrons. In contrast, insertion of impurities with electronic configurations similar to the host compound
does not reduce the conductance of the device. A full impurity interlayer, drastically suppresses
transmission via directed $d$ bonds, while $T(E_F,V=0)$ again is hardly altered.

\vspace*{-0.3cm}
\subsection*{Acknowledgement}
\vspace*{-0.3cm}
We gratefully acknowledge discussions with P.\ Schwab, U.\ Eckern, A.\ R.\ Rocha, and I.\ Rungger. The Au pseudopotential was provided by X.\ Lopez
(ETSF, Palaiseau, France). We thank the Deutsche Forschungsgemeinschaft (SFB 484) and the Egyptian Missions System for financial support.
The Smeagol project (SS) is sponsored by the Science Foundation of Ireland.

\end{document}